%% file: isit11_final.tex
\documentclass[conference,10pt]{IEEEtran}
 \IEEEoverridecommandlockouts

\usepackage[cmex10]{amsmath}
\usepackage{amssymb}
\usepackage{epsfig,epsf,pstricks,pgf}
\usepackage{cite}

\usepackage{enumerate}

\usepackage{amsthm}
\usepackage{psfrag}

\newtheorem{theorem}{Theorem}

\newtheorem{claim}{Claim}

\newtheorem{prop}{Proposition}

\theoremstyle{definition}

\newcommand{\snr}{\textnormal{\footnotesize \fontfamily{phv}\selectfont SNR}}
\newcommand{\snrs}{\textnormal{\tiny \fontfamily{phv}\selectfont SNR}}

\newcommand{\gskip}{\vspace{-.2cm}} 

\newcommand{\bE}{\mathbb{E}} 
\newcommand{\bP}{\mathbb{P}} 
\newcommand{\bY}{\mathbf{Y}} 
\newcommand{\bX}{\mathbf{X}}
\newcommand{\bW}{\mathbf{W}}

\newcommand{\bx}{\mathbf{x}}

\newcommand{\bA}{\mathbf{A}}

\newcommand{\one}{\boldsymbol{1}} 
\newcommand{\rw}{\rightarrow}

\begin{document}

\title{On the Role of Diversity in Sparsity Estimation}
\author{
\IEEEauthorblockN{Galen Reeves and Michael Gastpar\IEEEauthorrefmark{2}\thanks{\IEEEauthorrefmark{2} Also with the School of Computer and Communication Sciences, EPFL, Lausanne, Switzerland.}}
\IEEEauthorblockA{Department of Electrical Engineering and Computer Sciences \\
University of California, Berkeley }
}

\maketitle

\begin{abstract}
A major challenge in sparsity pattern estimation is that small modes are difficult to detect in the presence of noise. This problem is alleviated if one can observe samples from multiple realizations of the nonzero values for the same sparsity pattern. We will refer to this as ``diversity''. Diversity comes at a price, however, since each new realization adds new unknown nonzero values, thus increasing uncertainty. In this paper, upper and lower bounds on joint sparsity pattern estimation are derived. These bounds, which improve upon existing results even in the absence of diversity, illustrate key tradeoffs between the number of measurements, the accuracy of estimation, and the diversity. It is shown, for instance, that diversity introduces a tradeoff between the uncertainty in the noise and the uncertainty in the nonzero values. Moreover, it is shown that the optimal amount of diversity significantly improves the behavior of the estimation problem for both optimal and computationally efficient estimators. 

\end{abstract}

\section{Introduction}

An extensive amount of recent research in signal processing and statistics has focused on multivariate regression problems with sparsity constraints. One problem of particular interest, known as {\em sparsity pattern estimation}, is to determine which coefficients are nonzero using a limited number of observations. Remarkably, it has been shown that accurate estimation is possible using a relatively small number of (possibly noisy) linear measurements, provided that the number of nonzero values is relatively small (see e.g. \cite{Wainwright_InfoLimits_IEEE09,RG-Lower-Bounds,RG-Upper-Bounds}). 

It has also been shown that the presence of additional structure, beyond sparsity, can significantly alter the problem. Various examples include distributed or model-based compressed sensing \cite{BDWSB05, BarCevDuaHeg,VasWei10}, estimation from multiple measurement vectors \cite{Cotter2005}, simultaneous sparse approximation\cite{TroGilStr2006}, model selection \cite{YuaLin2006}, union support recovery \cite{Obozinski08}, multi-task learning \cite{Lounici09}, and estimation of block-sparse signals \cite{EldKupBol_IEEE2010,StoParHas2009}. 

In the present paper, we consider a joint sparsity pattern estimation framework motivated in part by the following engineering problem. Suppose that one wishes to estimate the sparsity patten of an unknown vector and is allowed to take either $M$ noisy linear measurements of the vector itself, or spread the same number measurements amongst multiple vectors with same sparsity pattern as the original vector, but different nonzero values. This type of problem arises, for example, in magnetic resonance imaging where the vectors correspond to images of the same body part (common sparsity pattern) viewed with different contrasting agents (different nonzero values). 

\begin{figure}
\psfrag{S}[B]{ x}
\psfrag{X}[B]{\small $\rho_\text{total}$} 
 \graphicspath{{inkscape-figs/}}
  \def\svgwidth{\columnwidth}
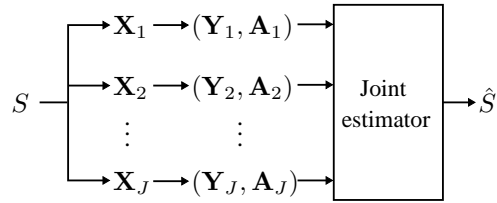
\caption{\label{fig:estimation_joint} Illustration of joint sparsity pattern estimation. The vectors $\bX_j$ share a common sparsity pattern $S$ but have independent nonzero values. The sparsity pattern $S$ is estimated jointly using measurements vectors $\bY_i$ corresponding to different measurement matrices $\bA_j$. }

\vspace{-.5cm}

\end{figure}

On one hand, splitting measurements across different vectors increases the number of unknown values, potentially making estimation more difficult. On the other hand, using all measurements on a single vector has the risk that nonzero values with small magnitudes will not be detected. To understand this tradeoff, this paper bounds the accuracy of various estimators for the estimation problem illustrated in Figure~\ref{fig:estimation_joint}. We refer to the number of vectors $J$ as the ``diversity''. 

\subsection{Overview of Contributions}


Several key contributions of this paper are the following:

\begin{itemize}
\item Our analysis improves upon previous work in the single-vector setting \cite{RG-Lower-Bounds,RG-Upper-Bounds} and shows that there exists a sharp divide between knowing almost everything and knowing almost nothing about the sparsity pattern, even in problem regimes where exact recovery is impossible.
\item Our bounds are relatively tight for a large range of problem parameters. Unlike bounds based on the restricted isometry property, they apply even when the number of measurements is small relative to the size of the sparsity pattern.

\item We show that the right amount of diversity is beneficial, but too much or too little can be detrimental (when the total number of measurements is fixed). Moreover, we show that diversity can significantly reduce the gap in performance between computationally efficient estimators, such the matched filter or LASSO, and estimators without any computational constraints. 

\end{itemize}







The remainder of the paper is outlined as follows. Theorem~1 gives a sufficient condition for a combinatorial estimator. Theorem~2 gives an information-theoretic necessary condition for any estimator. Theorem~3 gives a necessary and sufficient condition for a two-stage estimation architecture corresponding to either the matched filter, the LASSO, or the MMSE vector estimators, and Theorems 4-6 characterize various tradeoffs between the diversity, the number of measurements, the SNR, and the accuracy of estimation. 

Finally, we note that the joint estimation problem in this paper is closely related to the multiple measurement vector problem \cite{Cotter2005}, except that each vector is measured using a different matrix. 
Alternatively, our problem is a special case of block-sparsity \cite{EldKupBol_IEEE2010,StoParHas2009} with a block-sparse measurement matrix. 
Versions of our bounds for block-sparsity with dense measurement matrices can also be derived.

\subsection{Problem Formulation}

Let $\bX_1,\bX_2,\cdots,\bX_J \in \mathbb{R}^n$ be a set of jointly random sparse vectors whose nonzero values are indexed by a common sparsity pattern $S$
\begin{align}
S = \{ i : X_j(i) \ne 0 \}, \quad \text{for $j=1,2,\cdots,J$}.
\end{align}
We assume that $S$ is distributed uniformly over all subsets of $\{1,2,\cdots, n\}$ of size $k$ where $k$ is known, and that the nonzero values are i.i.d.~$\mathcal{N}(0,1)$\footnote{The results in this paper extend to any i.i.d.~distribution with bounded second moment. Due to space constraints, only the Gaussian case is presented.}.

We consider estimation of $S$ from measurement vectors $\bY_1,\bY_2,\cdots,\bY_J \in\mathbb{R}^m$ of the form 
\begin{align}
\bY_j = \sqrt{\textstyle \frac{\snrs}{k}}\, \bA_j \bX_j + \bW_j \quad \text{for $j=1,2,\cdots,J$}
\end{align}
where each $\bA_j\in \mathbb{R}^{m \times n}$ is a known matrix whose elements are i.i.d.~$\mathcal{N}(0,1)$ and $\bW_j\sim \mathcal{N}(0,I_{m \times m})$ is unknown noise. The estimation problem is depicted in Figure~\ref{fig:estimation_joint}. The accuracy of an estimate $\hat{S}$ is assessed using the (normalized) distortion function
\begin{align}\label{eq:distortion}
d(S,\hat{S}) =  \textstyle \frac{1}{k} \max\big( |S \backslash \hat{S}|,|\hat{S} \backslash S| \big)
\end{align}
where $|S \backslash \hat{S}|$ and $|\hat{S} \backslash S|$ denote the number of missed detections and false alarms respectively.

Our analysis considers the high dimensional setting where the diversity $J$ is fixed but the vector length $n$, sparsity $k$, and number of measurements per vector $m$ tend to infinity. We focus exclusively on the setting of linear sparsity where $k/n \rw \kappa$ for some fixed {\em sparsity rate} $\kappa \in (0,1/2)$ and $m/n \rw r$ for some fixed per-vector {\em sampling rate} $r >0$. The total number of measurements is given by $M= mJ$, and we use $ \rho = J r$ to denote the total sampling rate. We say that a {distortion} $\alpha \ge 0$ is {\em achievable} for an estimator $\hat{S}$ if $\Pr[ d(S, \hat{S}) > \alpha] \rw 0$ as $n \rw \infty$. 
The case $\alpha =0$ corresponds to exact recovery and the case $\alpha >0$ corresponds to a constant fraction of errors. 

\subsection{Notations}
For a matrix $A$ and set of integers $S$ we use $A(S)$ to denote the matrix formed by concatenating the columns of $A$ indexed by $S$. We use $H_b(p) = -p\log p - (1-p)\log(1-p)$ to denote binary entropy and all logarithms are natural.

\section{Joint Estimation Bounds}\label{sec:est_joint}

This section gives necessary and sufficient conditions for the joint sparsity pattern estimation problem depicted in Figure~\ref{fig:estimation_joint}.

One important property of the estimation problem is the relative size of the smallest nonzero values, averaged across realizations. For a given fraction $\beta \in [0,1]$, 
we define random variable
\vspace{-.2cm}
\begin{align}
P^{(n)}_J(\beta) = \arg\min_{\Delta \subset S\; :\;|\Delta| = \alpha k} \frac{1}{J} \sum_{j=1}^J \| \bX_j(\Delta)\|^2.
\end{align}
By the Glivenko-Cantelli theorem, $P_J^{(n)}(\beta)$ converges almost surely to a nonrandom limit $P_J(\beta)$. We will refer to this limit as the {\em diversity power}. If the nonzero values are Gaussian, as is assumed in this paper, it can be shown that
 \gskip
\begin{align}\label{eq:P_J}
P_J(\beta) =  \textstyle \int_0^\alpha \xi_J(p) dp
\end{align}
 \gskip
where 
 \gskip
\begin{align}\label{eq:xi}
\xi_J(p) =  \textstyle \big\{ t : \bP[ \frac{1}{J} \chi^2_J \le t ] = p\big\}
\end{align}
denotes the quantile function of a normalized chi-square random variable with $J$ degrees of freedom. 

Another important property is the metric entropy rate (in nats per vector length) of $S$ with respect to our distortion function $d(S,\hat{S})$. In \cite{RG-Lower-Bounds}, it is shown that this rate is given by
\begin{align}\label{eq:R_k}
R(\kappa,\alpha)=
{\textstyle H(\kappa ) - \kappa  H_b(\alpha) - (1\!-\!\kappa) H_b(\frac{\kappa \alpha}{1-\kappa})}
\end{align}
for all $\alpha < 1-\kappa$ and is equal to zero otherwise.

\subsection{Nearest Subspace Upper Bound}

We first consider the {\em nearest subspace} (NS) estimator which is given by
 \gskip
  \gskip
\begin{align}
\hat{S}^\text{\normalfont NS} = \arg\min_{S \; :\; |S|=k} \; \sum_{j=1}^J \text{\normalfont dist}(\bY_j , \bA_j(S))^2
\end{align}
where $\text{\normalfont dist}(\bY_j,\bA_j(S))$ denotes the euclidean distance between $\bY_j$ and the linear subspace spanned by the columns of $\bA_j(S)$. (For the case $J=1$, this estimator is known variously throughout the literature as $\ell_0$ minimization or maximum likelihood estimation.) 


\begin{theorem}\label{thm:NS_UB} For a given set $(\kappa,\snr, \rho,J)$, a distortion $\alpha$ is achievable for the nearest subspace estimator if
 \gskip
\begin{align}\label{eq:NS_UB}
\rho >   \kappa J + \max_{\beta \in [\alpha,1]} \min\big( E_1(\beta), E_2(\beta)\big)
\end{align}
 \gskip
where
\begin{align}
E_1(\beta)&= \frac{2H_b(\kappa) -  2R(\kappa,\beta)  + 2 \beta \kappa J  \log(5/3)}{\frac{1}{J} \log \big( 1+ \frac{4}{25} J P_J(\beta) \,\snr  \big)} \\
E_2(\beta) 
&= \frac{2 H_b(\kappa) - 2R(\kappa,\beta)}{\log\big( 1+ P_1(\beta)\, \snr\big)  + 1/\big(P_1(\beta) \, \snr\big) -1}
\end{align}
with 
$P_J(\cdot)$ given by \eqref{eq:P_J} and $R(\cdot,\cdot)$ given by \eqref{eq:R_k}. 
\end{theorem}




Theorem~\ref{thm:NS_UB} is a combination of two bounds. The part due to $E_1(\beta)$ determines the scaling behavior at low distortions and low SNR and the part due to $E_2(\beta)$ determines the scaling behavior at high SNR.
One important property of $E_1(\beta)$ is that its denominator scales linearly with the effective power of the $P_J(\beta) \, \snr$ when when $\beta$ is small. As a consequence, Theorem~\ref{thm:NS_UB}  closes a gap in previous bounds for the case $J=1$ and correctly characterizes the boost in performance due to the diversity when $J>1$.

\subsection{Optimal Estimation}

We next consider an information-theoretic lower bound on the distortion for any estimator. This bound depends on the entropy of the smallest nonzero values. For a given fraction $\beta \in [0,1]$, we define the {\em conditional entropy power} 
\begin{align}
\mathcal{N}(\beta)= \textstyle \frac{1}{2 \pi e} \exp \big\{-2 h\big(U | U^2 \le \xi_1(\beta)\big)\big\}
\end{align}
where $h(\cdot)$ is differential entropy and $U \sim \mathcal{N}(0,1)$.

\begin{theorem}\label{thm:LB}For a given set $(\kappa,\snr, \rho,J)$, a distortion $\alpha$ is not achievable for any estimator if
\begin{align}\label{eq:LB2}
\max_{\beta \in [0,1]} \Big\{& R\big({\textstyle \frac{\beta\kappa}{1-\kappa+\beta\kappa}},{\textstyle \frac{\alpha}{\beta}} \big) -J \min\big(  \Lambda_1(\beta),\Lambda_2(\beta)\big) \Big\} >0
\end{align}
where
\gskip
\gskip
\begin{align}
\Lambda_1(\beta) &=  \mathcal{V}_1\big({\textstyle \frac{\rho}{1-\kappa+\beta\kappa}},P_J^2(\beta) \, \snr\big) \\
\Lambda_2(\beta) &=\mathcal{V}_1\big({\textstyle \frac{\rho}{1-\kappa+\beta\kappa}},\beta^{1-1/J} P_{1}(\beta^{1/J}) \,\snr\big) \nonumber  \\
& \quad - {\textstyle \frac{\beta\kappa}{1-\kappa+\beta\kappa}} \mathcal{V}_2\big({\textstyle \frac{\rho}{\beta \kappa}},\beta\, \mathcal{N}(\beta^{1/J}) \, \snr \big)
\end{align}
with
\gskip
\gskip
\begin{align}\label{eq:infoGeneral}
\mathcal{V}_1(r,\gamma)&=
\begin{cases}
\frac{r}{2} \log(1+\gamma),& \text{\normalfont if $r\le 1$}\\
\frac{1}{2} \log(1+ r \gamma),&\text{\normalfont if $ r >1$}
\end{cases}\\
\mathcal{V}_2(r,\gamma)&=
\begin{cases}
\frac{r}{2} \log\big(1+\gamma \Delta(r) \big ),& \text{\normalfont if $r < 1$}\\
\frac{1}{2} \log\big(1+ r \gamma\Delta(\frac{1}{r}) \big ),&\text{\normalfont if $ r >1$}
\end{cases}
\end{align}
and $\Delta(r) = e^{-1}(1-r)^{1-1/r}$.
\end{theorem}

Theorem~\ref{thm:LB} is also a combination of two bounds. The part due to $\Lambda_1(\beta)$ determines the scaling behavior at low distortions and low SNR and the part due to $\Lambda_2(\beta)$ determines the scaling behavior at high SNR. As was the case for the nearest subspace upper bound, this bound is inversely proportional to the effective power $P_J(\beta) \, \snr$ when the effective power is small.

\section{Two-Stage Estimation Bounds}\label{sec:est_dist}
\begin{figure}
\psfrag{S}[B]{ x}
\psfrag{X}[B]{\small $\rho_\text{total}$} 
 \graphicspath{{inkscape-figs/}}
  \def\svgwidth{\columnwidth}
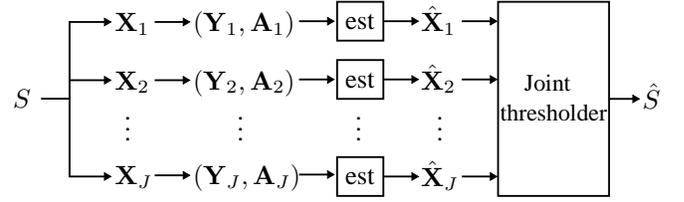
\caption{\label{fig:estimation_dist}Illustration of single-vector estimation followed by joint thresholding.}

\vspace{-.2cm}

\end{figure}

This section gives bounds for the two-stage estimation architecture depicted in Figure \ref{fig:estimation_dist}. In the first stage, each vector $\bX_j$ is estimated from its measurements $\bY_j$. In the second stage, the sparsity pattern $S$ is estimated by jointly thresholding estimates $\hat{\bX}_1, \hat{\bX}_2,\cdots,\hat{\bX}_J$. One advantage of this architecture is that the estimation in the first stage can be done in parallel. We will see that this architecture can be near optimal in some settings but is highly suboptimal in others.

\subsection{Single-Vector Estimation}

Three different estimators are considered: the {\em matched filter} (MF), the {\em LASSO}, and the {\em minimum mean squared error} estimator (MMSE). Recent results have shown that the asymptotic behavior of these estimators can be characterized in terms of an equivalent scalar estimation problem. Since these results correspond to the case $J=1$, we use the notation $\bX$ and $\bY$ and use the per-vector sampling rate $r$ instead of the total sampling rate $\rho$. Also, we define the sparse Gaussian distribution 
\begin{align}
F_\kappa(x) =  \textstyle \kappa \int_{-\infty}^x \frac{1}{\sqrt{2 \pi}} e^{-\frac{u^2}{2}}du  + (1-\kappa) \one(x \le 0)
\end{align}
which corresponds to the marginal distribution of $X(i)$. 

The first result characterizes the asymptotic behavior of the matched filter which is given by
\begin{align}\label{eq:XhatMF}
\hat{\bX}^\text{\normalfont MF} =  \textstyle \frac{1}{m} \sqrt{\frac{k}{\snrs}}  \bA^T \bY.
\end{align}
To our knowledge, this result was first shown (with convergence in probability) in \cite{RG09,RG-Upper-Bounds}. Almost sure convergence follows from recent tools developed in \cite{BM10a}.

\begin{prop}[Matched Filter]
\label{thm:EDFmatchedfilter} 
The empirical distribution on the elements of $(\bX,\hat{\bX}^\text{\normalfont MF})$ converges weakly and almost surely to the distribution on $(X,X+ \sigma W)$ where $X\sim F_\kappa$ and $W\sim \mathcal{N}(0,1)$ are independent and
\begin{align}\label{eq:sig2_MF}
\sigma^2 = \frac{\kappa}{r} \Big[\frac{1}{\snr} +1\Big].
\end{align}
\end{prop}

The next result, due to Donoho et al. \cite{DMM09} and Bayati and Montanari,\cite{BM10a}, describes the asymptotic behavior of the LASSO which is given by
\begin{align}\label{eq:XhatLASSO}
\hat{\bX}^\text{LASSO} = \arg \inf_{\bx \in \mathbb{R}^n} {\textstyle \frac{1}{2} } \|\bY - \sqrt{\textstyle\frac{\snrs}{k} }\bA \bx\|^2_2 + \lambda \|\bx\|_1\end{align}
where $\lambda \ge 0$ is a regularization parameter.

\begin{prop}[{LASSO}]
\label{thm:EDFlasso}
The emprical distribution on the elements of $(\bX,\hat{\bX}^\text{\normalfont LASSO})$ converges weakly almost surely to the distribution on $(X,\eta_t(X +\sigma W))$ where 
$X \sim F_\kappa$ and $W \sim \mathcal{N}(0,1)$ are independent, $\eta_t(x) = [x - \text{\normalfont sign}(x)t] \one(|x| >t)$,  
and $\sigma^2$ and $t$ are given by the fixed point equations
\begin{align}
 \sigma^2 &=  \frac{1}{r} \left[\frac{\kappa}{\snr} + \bE\big[|X - \eta_t(X+\sigma W)|^2 \big] \right] \label{eq:sig2_LASSO}\\ 
 t & = \frac{1}{r} \left[ \frac{\kappa}{\snr} \lambda + t \Pr \big[|X+\sigma W| >t \big] \right]. \label{eq:t_LASSO}
\end{align}
\end{prop}

\begin{figure*}
\centering
\psfrag{t1}[]{\footnotesize Diversity $J=1$}
\psfrag{x1}[]{\footnotesize $\snr$ (dB)}
\psfrag{y1}[c][c][1][0]{\footnotesize total sampling rate $\rho$}
\psfrag{title}{}
\includegraphics[width=.31 \textwidth, trim = 0 .5cm 0 1cm, clip]{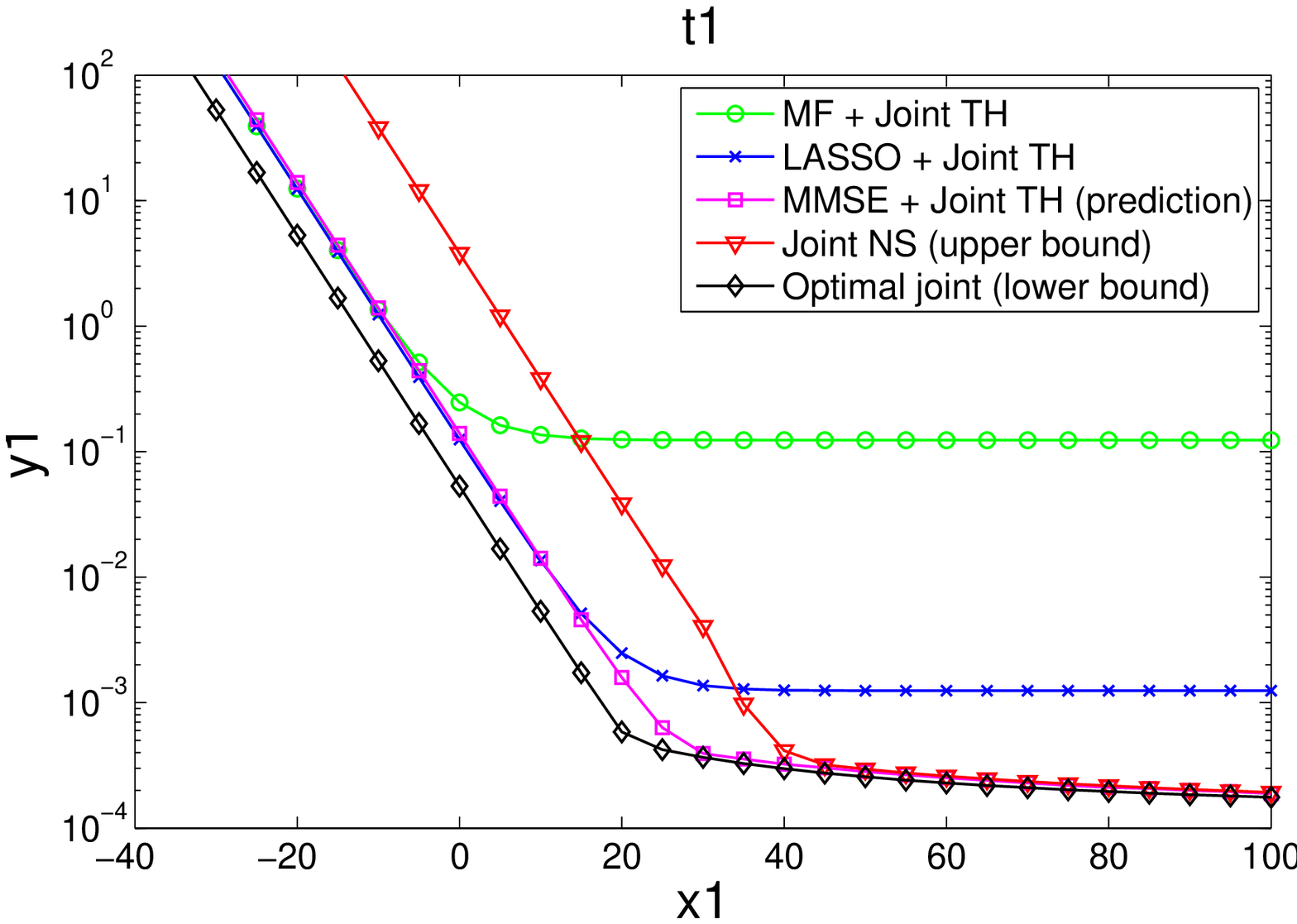}
\psfrag{t1}[]{\footnotesize Diversity $J=4$}
\includegraphics[width=.31 \textwidth, trim = 0 .5cm 0 1cm, clip]{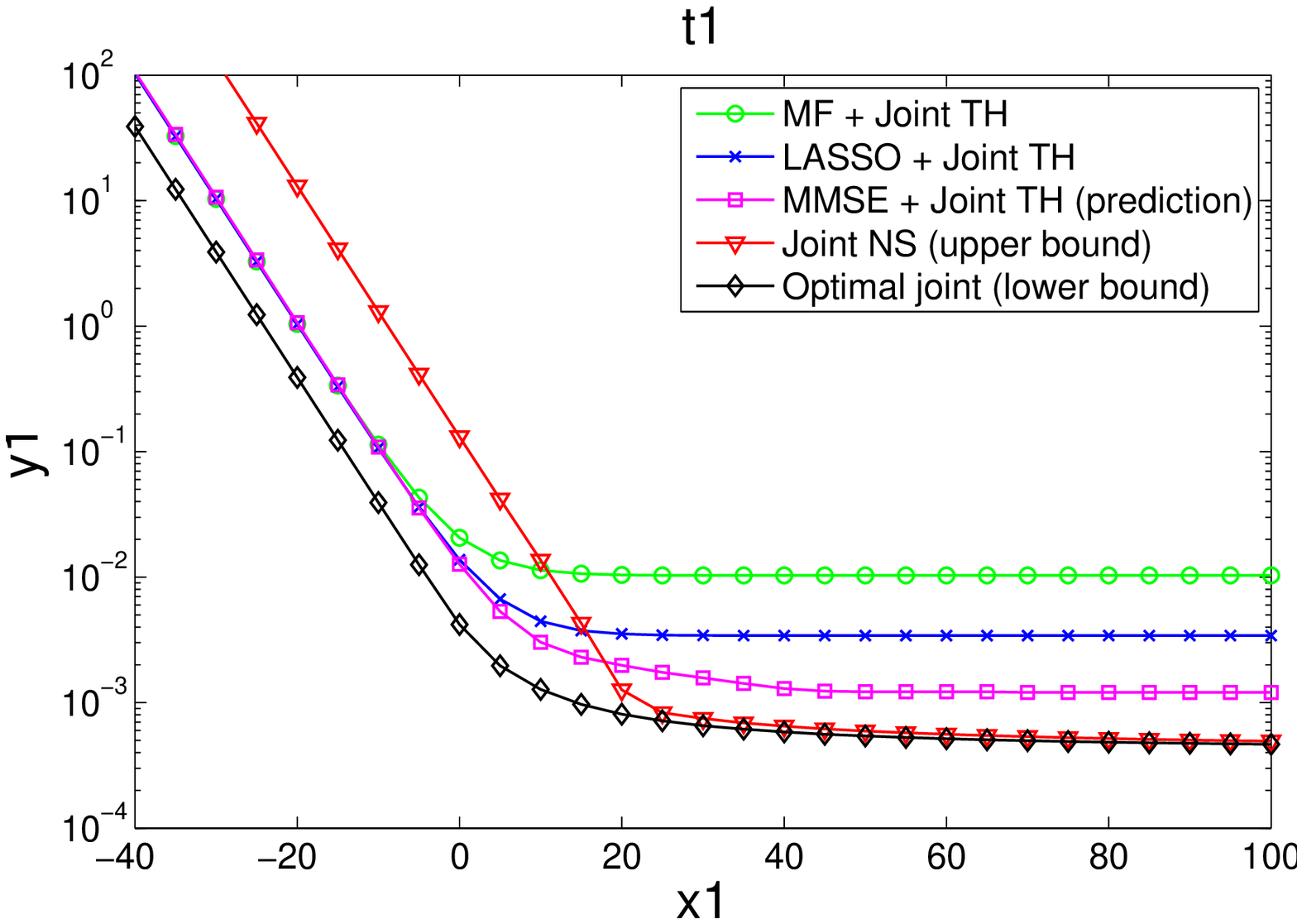}
\psfrag{t1}[]{\footnotesize Diversity $J=16$}
\includegraphics[width=.31 \textwidth, trim = 0 .5cm 0 1cm, clip]{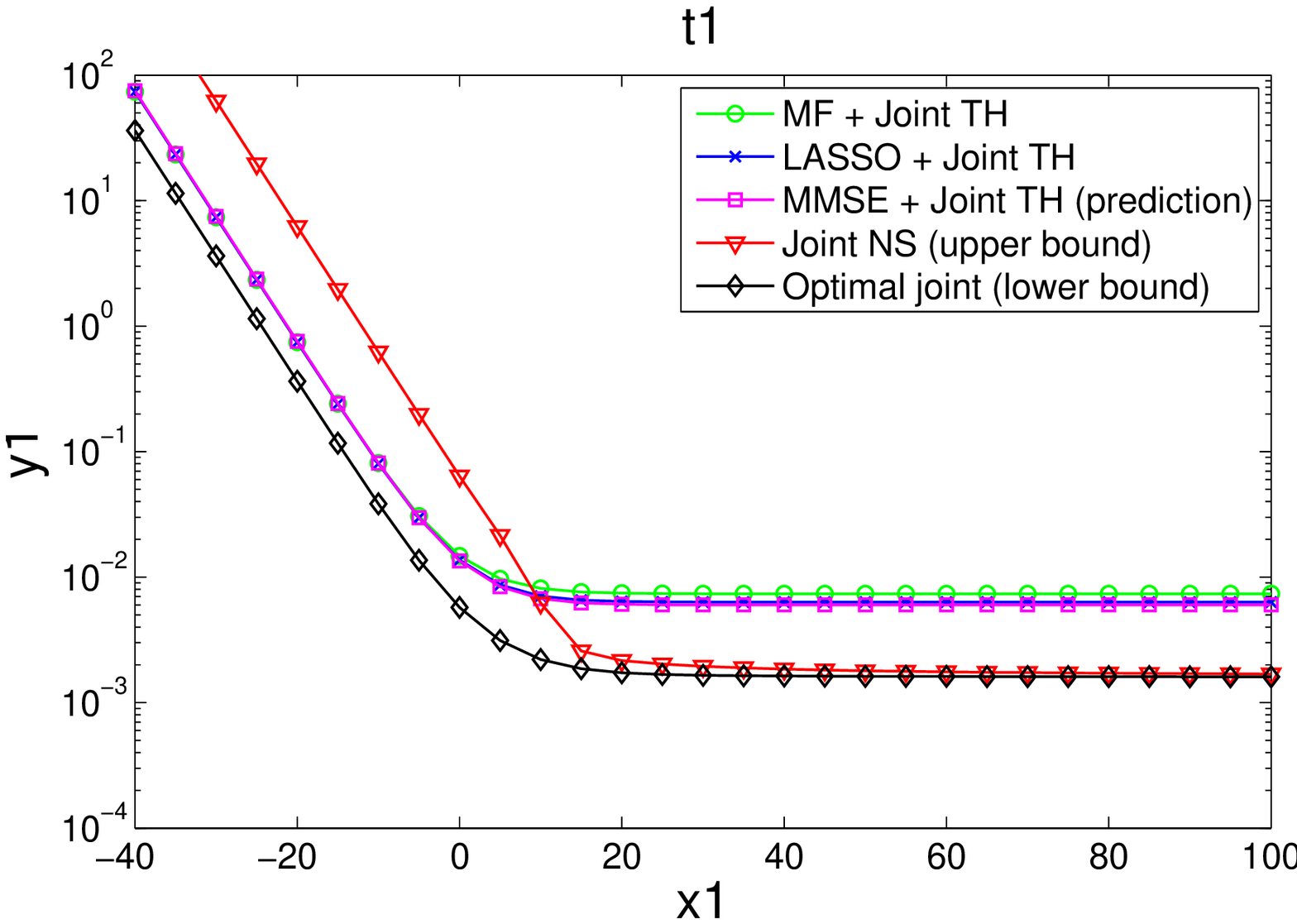}
\caption{\label{fig:RvSNR}Bounds on the total sampling rate $\rho = J r$ as a function of $\snr$ for various $J$ when $\alpha = 0.1$ and $\kappa = 10^{-4}$.}


\end{figure*}

\begin{figure*}
\centering
\psfrag{t1}[]{\footnotesize Diversity $J=1$}
\psfrag{x1}[]{\footnotesize total sampling rate $\rho$}
\psfrag{y1}[c][c][1][0]{\footnotesize distortion $\alpha$}
\psfrag{title}{}
\includegraphics[width=.31 \textwidth, trim = 0 .5cm 0 1cm, clip]{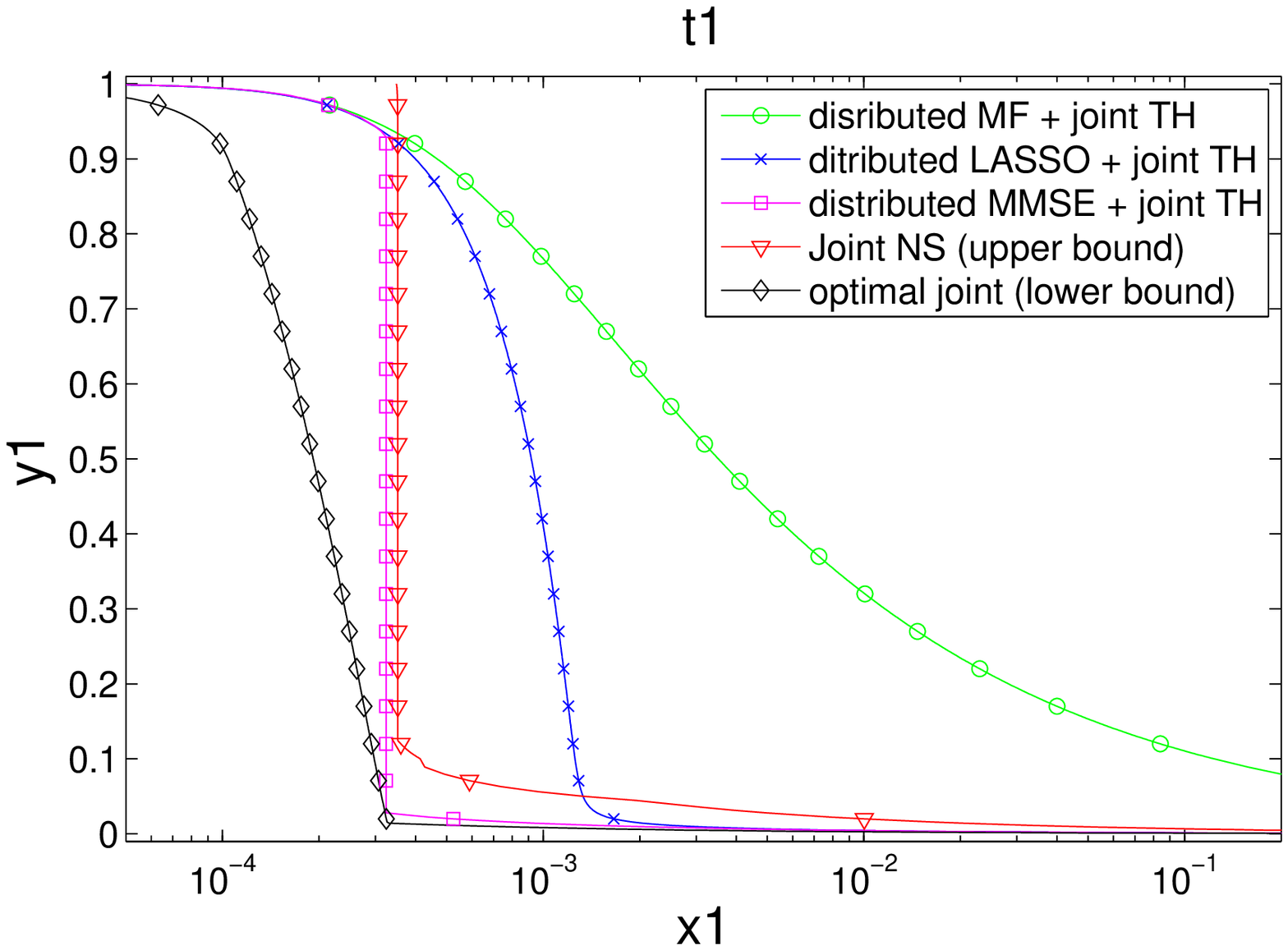}
\psfrag{t1}[]{\footnotesize Diversity $J=4$}
\includegraphics[width=.31 \textwidth, trim = 0 .5cm 0 1cm, clip]{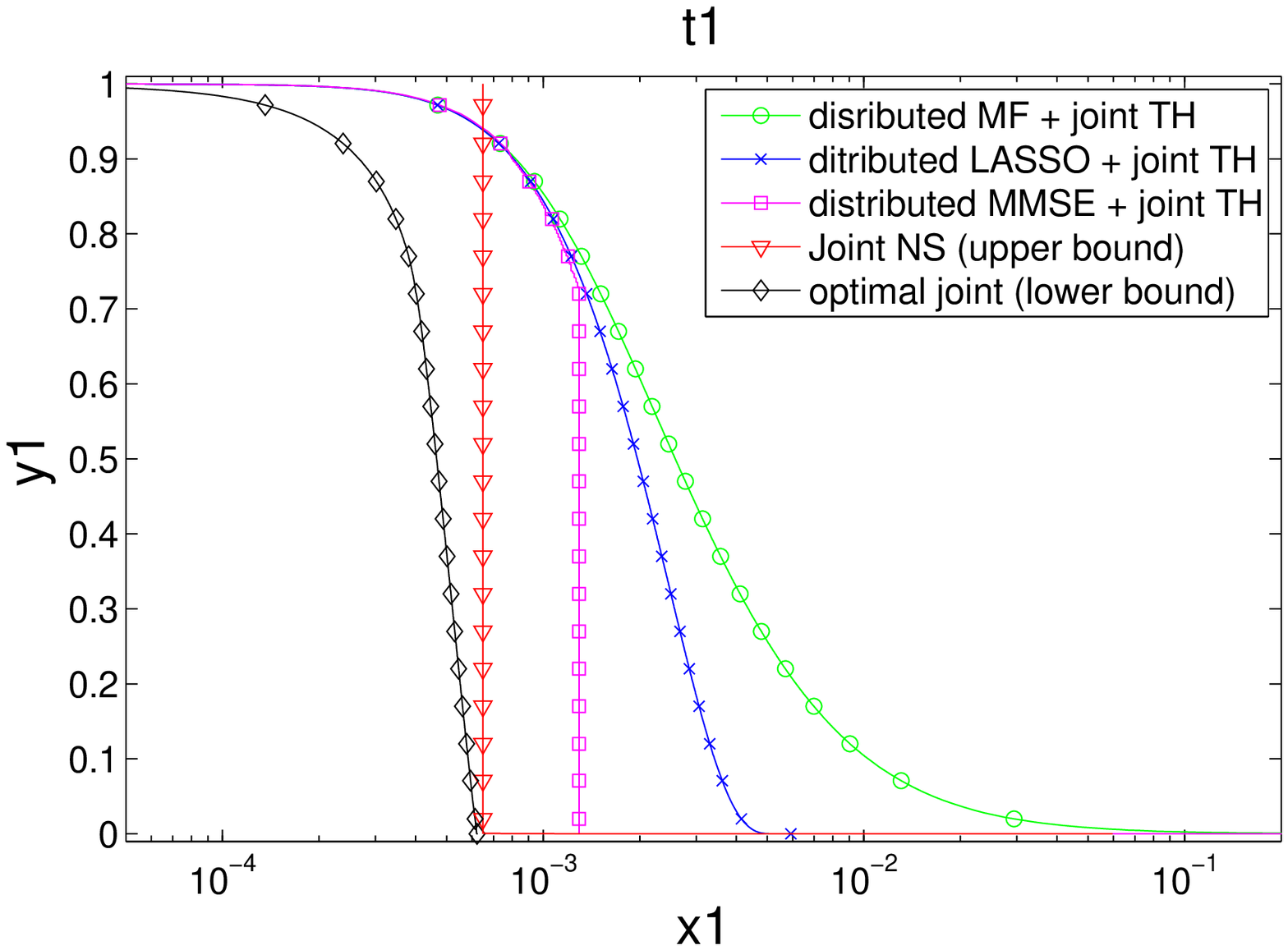}
\psfrag{t1}[]{\footnotesize Diversity $J=16$}
\includegraphics[width=.31 \textwidth, trim = 0 .5cm 0 1cm, clip]{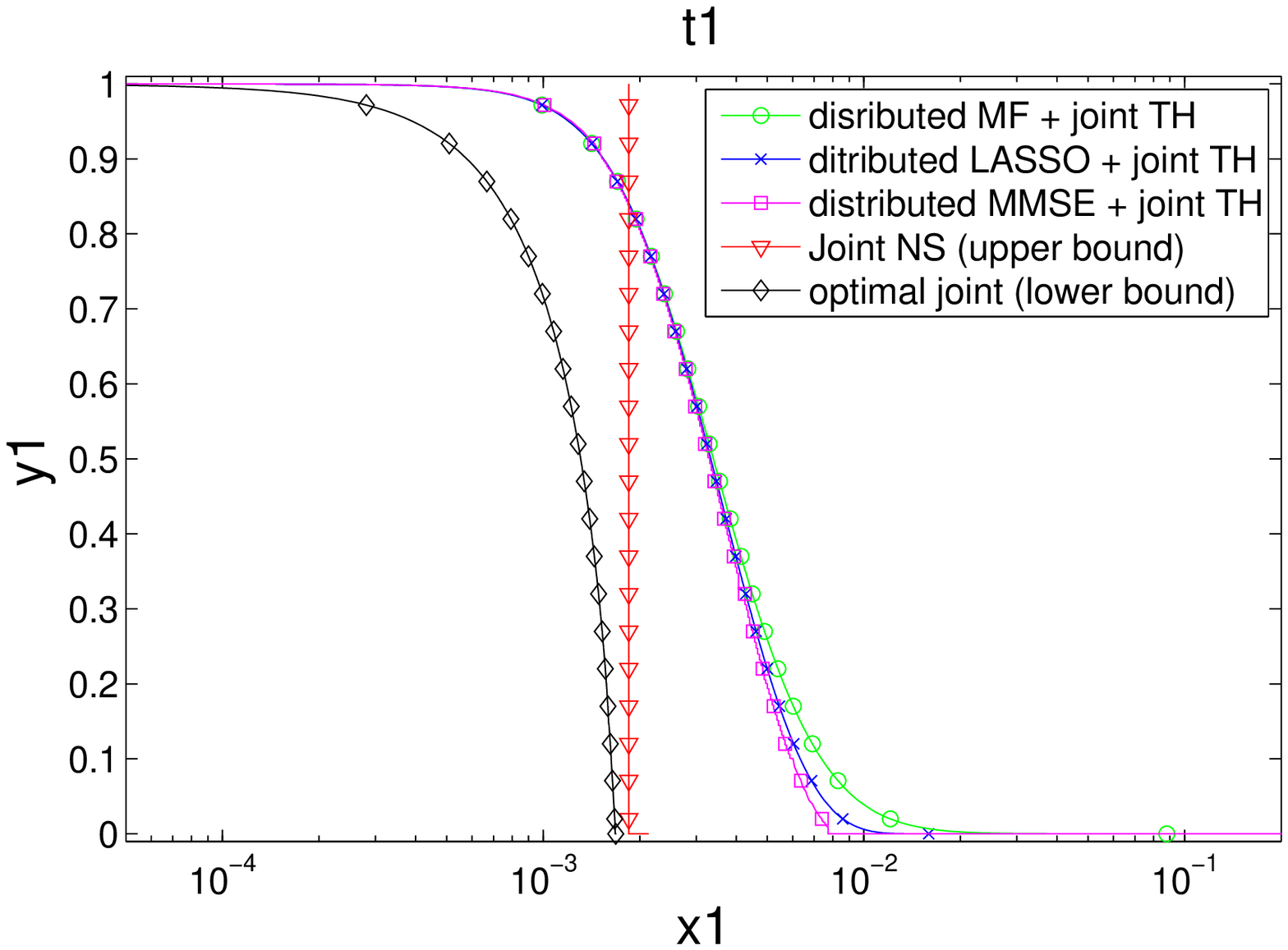}
\caption{\label{fig:DvR}Bounds on the distortion $\alpha$ as a function of the total sampling rate $\rho = Jr$ for various $J$ when $\snr = 40$ dB and $\kappa = 10^{-4}$.}

\vspace{-.5cm}

\end{figure*}

The final result, based on the work of Guo and Verdu \cite{GV05}, characterizes the asymptotic behavior of the MMSE which is given by
\begin{align}
\hat{\bX}^\text{\normalfont MMSE} = \bE[ \bX | \bY].
\end{align}
This result depends on a powerful but non-rigorous replica method, and is thus stated as a claim. 

\begin{claim}[MMSE]
\label{thm:mmse}
The distribution on the elements of $(\bX,\hat{\bX}^\text{\normalfont MMSE})$ converges weakly in expectation to the distribution on $(X,\bE[X | X+\sigma W])$ where 
$X \sim F_\kappa$ and $W \sim \mathcal{N}(0,1)$ are independent and $\sigma^2$ is given by
\begin{align}\label{eq:sig2_MMSE}
\sigma^2 = \arg \min_{\sigma^2 \ge 0} \big\{ r \log \sigma^2  + \frac{\kappa}{\snr\, \sigma^2} + 2 \, I(X; X+\sigma W) \big \}.
\end{align}
\end{claim}

\subsection{Thresholding}

For the second stage of estimation we consider the {\em joint thresholding} sparsity pattern estimator given by
\begin{align}\label{eq:S_joint_thresholding}
\textstyle \hat{S}^\text{\normalfont TH}=  \big \{ i :  \sum_{j=1}^J \hat{X}^2_j(i) \ge t \big \}
\end{align}
where the threshold $t\ge 0$ is chosen to minimize the expected distortion. Since this estimator evaluates each index $i \in \{1,2,\cdots,n\}$ independently, and since the estimated vectors $\bX_1,\bX_2,\cdots,\bX_j$ are conditionally independent given the sparsity pattern $S$, the distribution on the distortion $d(S,\hat{S}^\text{TH})$ can be characterized by joint distribution on $(X_1(1),\hat{X}_1(1))$. 


The following result describes the relationship between the distortion $\alpha$, the diversity $J$, and the effective noise power $\sigma^2$. 

\begin{theorem}\label{thm:TH}
Suppose that for $j =1,2,\cdots, J$, the empirical joint distributions on the elements of $(\bX_j,\hat{\bX}_j)$ converge weakly to one of the scalar distributions corresponding to the matched filter (Proposition~\ref{thm:EDFmatchedfilter}),  the LASSO (Proposition~\ref{thm:EDFlasso}), or the MMSE (Claim~\ref{thm:mmse}) with noise power $\sigma^2$. Then, a distortion $\alpha$ is achievable for the thresholding estimator \eqref{eq:S_joint_thresholding} if and only if $\sigma^2 \ge \sigma^2_J(\alpha)$ where
\vspace{-.2cm}
\begin{align}\label{eq:sigma2_alpha}
\sigma_J^2(\alpha) =  \frac{\xi_{J}(\alpha)}{\xi_J(1 - \frac{\alpha \kappa}{1-\kappa}) -\xi_{J}(\alpha) }
\end{align}
with $\xi_J(\alpha)$ given by \eqref{eq:xi}.
\end{theorem}


Theorem~\ref{thm:TH} shows that the relationship between $\alpha$ and $J$ is encapsulated by the term $\sigma^2_J(\alpha)$. With at bit of work it can be shown that the numerator and denominator in \eqref{eq:sigma2_alpha} scale like $\alpha^{-1}P_J(\alpha)$ and $\alpha^{-1}R(\kappa,\alpha)$ respectively when $\alpha$ is small. 
Thus, plugging $\sigma_J^2(\alpha)$ into the equivalent noise expression of the matched filter given in \eqref{eq:sig2_MF} shows that bounds attained using Theorem~\ref{thm:TH} have similar low distortion behavior to the bounds in Section~\ref{sec:est_joint}.

One advantageous property of Theorem~\ref{thm:TH} is that the bounds are exact. As a consequence, these bounds are sometimes lower than the upper bound in Theorem~\ref{thm:NS_UB}, which is loose in general. One shortcoming however, is that the two-stage architecture does not take full advantage of the joint structure during the first stage of estimation. As a consequence, the performance of these estimators can be highly suboptimal, especially at high SNR.

\section{Sampling - Diversity Tradeoff}

In this section, we analyze various behaviors of the bounds in Theorems 1, 2, and 3, with an emphasis on the tradeoff provided by the diversity $J$. The following results characterize the high SNR and low distortion behavior of optimal estimation. 

\begin{theorem}[High SNR]\label{thm:highSNR}
Let $(\kappa,J,\alpha)$, be fixed and let $\rho(\snr)$ denote the infimum over sampling rates $\rho$ such that $\alpha$ is achievable for the optimal estimator. Fix any $\epsilon >0$. 
\begin{enumerate}[(a)]
\item If $\alpha >0$, then
\vspace{-.1cm}
\begin{align}
\rho(\snr) \le J \kappa + \frac{2 H_b(\kappa)(1+\epsilon)}{\log \snr} 
\end{align}
for all $\snr$ large enough.
\item If $2 R(\kappa,\alpha) > J \kappa$, then
\vspace{-.1cm}
\begin{align}
\rho(\snr) \ge J \kappa + \frac{2 R(\kappa,\alpha)(1-\epsilon)}{\log \snr} 
\end{align}
for all $\snr$ large enough.
\end{enumerate}
\end{theorem}

\begin{theorem}[Low Distortion]\label{thm:lowD}
Let $(\kappa,J,\snr)$ be fixed and let $\rho(\alpha)$ denote the infimum over sampling rates $\rho$ such that $\alpha$ is achievable for the optimal estimator. There exist constants $0 < C^- \le C^+ < \infty$ such that
\begin{align}
C^- \textstyle \big(\frac{1}{\alpha}\big)^{2/J} \log(\frac{1}{\alpha}\big) \le \rho(\alpha) \le C^+ \textstyle \big(\frac{1}{\alpha}\big)^{2/J} \log(\frac{1}{\alpha}\big)
\end{align}
for all $\alpha$ small enough. 
\end{theorem}

Theorems~\ref{thm:highSNR} and \ref{thm:lowD} illustrate a tradeoff. At high SNR, the difficulty of estimation is dominated by the uncertainty about the nonzero values. Accordingly, the number of measurements is minimized by letting $J=1$. As the desired distortion becomes small however, the opposite behavior occurs. Since estimation is limited by the size of the smallest nonzero values, it is optimal to choose $J$ large to increase the diversity power. This behavior can be seen, for example, in Figures 3-6.

A natural question then, is how does one best choose the diversity $J$? The following result shows that the right amount of diversity can significantly improve performance.

\begin{theorem}\label{thm:optJ}
Let $(\kappa,\snr)$ be fixed and let $\rho(\alpha,J)$ denote the infimum over sampling rates $\rho$ such that $\alpha$ is achievable with diversity $J$. Then, 
\vspace{-.1cm}
\begin{align}
\rho(\alpha,J) \le \kappa J + \textstyle O\big(\frac{\alpha}{P(\alpha,J)}\big).
\end{align}
Moreover, if $J =J^*(\alpha) =  \Theta(\log(1/\alpha)$ then
\vspace{-.1cm}
\begin{align}
\rho(\alpha,J^*(\alpha)) = \Theta( \log(1/\alpha)).
\end{align}
\end{theorem}

An important implication of Theorem~\ref{thm:optJ} is that the optimal choice of $J$ allows the distortion to decay {\em exponentially} rapidly with the sampling rate $\rho$. Note that the rate of decay is only polynomial if $J$ is fixed. Interestingly, it can also be shown that the same exponential boost can be obtained using non-optimal estimators, albeit with smaller constants in the exponent.  

\begin{figure}
\centering
\psfrag{t1}[]{}
\psfrag{x1}[]{\footnotesize total sampling rate $\rho$}
\psfrag{y1}[c][c][1][0]{\footnotesize distortion $\alpha$}
\psfrag{title}{}
\includegraphics[width=0.8 \columnwidth, trim = 0 .5cm 0 1cm, clip]{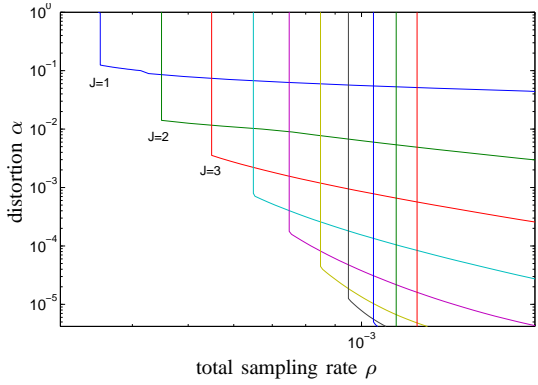}
\caption{\label{fig:LowD_NS} The upper bound (Theorem~\ref{thm:NS_UB}) on the total sampling rate $\rho= J r$ of the nearest subspace estimator as a function of the distortion $\alpha$ for various $J$ when $\snr = 40$ dB and $\kappa = 10^{-4}$.}

\vspace{-.2cm}

\end{figure}

\begin{figure}
\centering
\psfrag{t1}[]{}
\psfrag{x1}[]{\footnotesize total sampling rate $\rho$}
\psfrag{y1}[c][c][1][0]{\footnotesize distortion $\alpha$}
\psfrag{title}{}
\includegraphics[width=0.8 \columnwidth, trim = 0 0.5cm 0 1cm, clip]{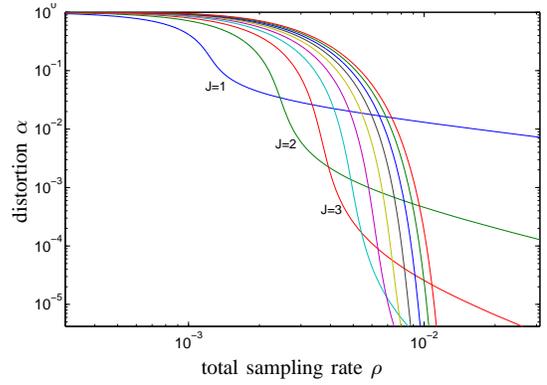}
\caption{\label{fig:LowD_LASSO} The upper bound (Theorem~\ref{thm:TH}) on the total sampling rate $\rho=J r$ of LASSO + Joint Thresholding as a function of the distortion $\alpha$ for various $J$ when $\snr = 30$ dB and $\kappa = 10^{-4}$.}

\vspace{-.5cm}

\end{figure}

The effect of the diversity $J$ is illustrated in Fig.~\ref{fig:LowD_NS} for the nearest subspace estimator and in Fig.~\ref{fig:LowD_LASSO} for Lasso + thresholding. In both cases, the bounds show the same qualitative behavior--each value of the diversity $J$ traces out a different curve in the sampling rate distortion region. It is important to note however, that due to the sub-optimality of the two stage architecture and the LASSO estimator, these similar behaviors occur only at different SNRs and with an order of magnitude difference in the sampling rate. 

\section*{Acknowledgment}
This work was supported in part by ARO MURI No. W911NF-06-1-0076, and in part by 3TU.CeDICT: Centre for Dependable ICT Systems, The Netherlands.

\bibliographystyle{ieeetran}
\bibliography{isit11_final.bbl}
\end{document}

%% file: inkscape-figs/estimation_joint.eps_tex

\begingroup
  \makeatletter
  \providecommand\color[2][]{%
    \errmessage{(Inkscape) Color is used for the text in Inkscape, but the package 'color.sty' is not loaded}
    \renewcommand\color[2][]{}%
  }
  \providecommand\transparent[1]{%
    \errmessage{(Inkscape) Transparency is used (non-zero) for the text in Inkscape, but the package 'transparent.sty' is not loaded}
    \renewcommand\transparent[1]{}%
  }
  \providecommand\rotatebox[2]{#2}
  \ifx\svgwidth\undefined
    \setlength{\unitlength}{288.8pt}
  \else
    \setlength{\unitlength}{\svgwidth}
  \fi
  \global\let\svgwidth\undefined
  \makeatother
  \begin{picture}(1,0.29518698)%
    \put(0,0){\includegraphics[width=\unitlength]{estimation_joint.eps}}%
    \put(0.09719334,0.13412614){\color[rgb]{0,0,0}\makebox(0,0)[lb]{\smash{$S$}}}%
    \put(0.24954798,0.24996081){\color[rgb]{0,0,0}\makebox(0,0)[lb]{\smash{$\bX_1$}}}%
    \put(0.24954798,0.15931751){\color[rgb]{0,0,0}\makebox(0,0)[lb]{\smash{$\bX_2$
}}}%
    \put(0.24954796,0.01604244){\color[rgb]{0,0,0}\makebox(0,0)[lb]{\smash{$\bX_J$}}}%
    \put(0.36704918,0.24996081){\color[rgb]{0,0,0}\makebox(0,0)[lb]{\smash{$(\bY_1,\bA_1)$}}}%
    \put(0.36704918,0.15931751){\color[rgb]{0,0,0}\makebox(0,0)[lb]{\smash{$(\bY_2,\bA_2)$}}}%
    \put(0.36704918,0.01604254){\color[rgb]{0,0,0}\makebox(0,0)[lb]{\smash{$(\bY_J, \bA_J)$}}}%
    \put(0.26616841,0.08329426){\color[rgb]{0,0,0}\makebox(0,0)[lb]{\smash{$\vdots$}}}%
    \put(0.43514347,0.08329426){\color[rgb]{0,0,0}\makebox(0,0)[lb]{\smash{$\vdots$}}}%
    \put(0.59303819,0.11099515){\color[rgb]{0,0,0}\makebox(0,0)[lb]{\smash{\small estimator}}}%
    \put(0.62073905,0.15808651){\color[rgb]{0,0,0}\makebox(0,0)[lb]{\smash{\small Joint}}}%
    \put(0.79802448,0.13300162){\color[rgb]{0,0,0}\makebox(0,0)[lb]{\smash{$\hat{S}$}}}%
  \end{picture}%
\endgroup

%% file: inkscape-figs/estimation_dist.eps_tex

\begingroup
  \makeatletter
  \providecommand\color[2][]{%
    \errmessage{(Inkscape) Color is used for the text in Inkscape, but the package 'color.sty' is not loaded}
    \renewcommand\color[2][]{}%
  }
  \providecommand\transparent[1]{%
    \errmessage{(Inkscape) Transparency is used (non-zero) for the text in Inkscape, but the package 'transparent.sty' is not loaded}
    \renewcommand\transparent[1]{}%
  }
  \providecommand\rotatebox[2]{#2}
  \ifx\svgwidth\undefined
    \setlength{\unitlength}{288.8pt}
  \else
    \setlength{\unitlength}{\svgwidth}
  \fi
  \global\let\svgwidth\undefined
  \makeatother
  \begin{picture}(1,0.2938885)%
    \put(0,0){\includegraphics[width=\unitlength]{estimation_dist.eps}}%
    \put(0.01409084,0.13366844){\color[rgb]{0,0,0}\makebox(0,0)[lb]{\smash{$S$}}}%
    \put(0.16644542,0.24862334){\color[rgb]{0,0,0}\makebox(0,0)[lb]{\smash{$\bX_1$}}}%
    \put(0.16644542,0.16136933){\color[rgb]{0,0,0}\makebox(0,0)[lb]{\smash{$\bX_2$
}}}%
    \put(0.16644542,0.01885433){\color[rgb]{0,0,0}\makebox(0,0)[lb]{\smash{$\bX_J$}}}%
    \put(0.28394654,0.24862334){\color[rgb]{0,0,0}\makebox(0,0)[lb]{\smash{$(\bY_1,\bA_1)$}}}%
    \put(0.28394654,0.16136933){\color[rgb]{0,0,0}\makebox(0,0)[lb]{\smash{$(\bY_2,\bA_2)$}}}%
    \put(0.28394656,0.01885443){\color[rgb]{0,0,0}\makebox(0,0)[lb]{\smash{$(\bY_J, \bA_J)$}}}%
    \put(0.1814535,0.08463064){\color[rgb]{0,0,0}\makebox(0,0)[lb]{\smash{$\vdots$}}}%
    \put(0.35042857,0.08463064){\color[rgb]{0,0,0}\makebox(0,0)[lb]{\smash{$\vdots$}}}%
    \put(0.74539277,0.11497715){\color[rgb]{0,0,0}\makebox(0,0)[lb]{\smash{\small thresholder}}}%
    \put(0.78140389,0.15929844){\color[rgb]{0,0,0}\makebox(0,0)[lb]{\smash{\small Joint}}}%
    \put(0.95868924,0.13228472){\color[rgb]{0,0,0}\makebox(0,0)[lb]{\smash{$\hat{S}$}}}%
    \put(0.5127058,0.25153168){\color[rgb]{0,0,0}\makebox(0,0)[lb]{\smash{est}}}%
    \put(0.6262792,0.24862334){\color[rgb]{0,0,0}\makebox(0,0)[lb]{\smash{$\hat{\bX}_1$}}}%
    \put(0.5127058,0.16427773){\color[rgb]{0,0,0}\makebox(0,0)[lb]{\smash{est}}}%
    \put(0.62627924,0.16136923){\color[rgb]{0,0,0}\makebox(0,0)[lb]{\smash{$\hat{\bX}_2$}}}%
    \put(0.5127058,0.01885449){\color[rgb]{0,0,0}\makebox(0,0)[lb]{\smash{est}}}%
    \put(0.6262792,0.01594601){\color[rgb]{0,0,0}\makebox(0,0)[lb]{\smash{$\hat{\bX}_J$}}}%
    \put(0.52771385,0.08463064){\color[rgb]{0,0,0}\makebox(0,0)[lb]{\smash{$\vdots$}}}%
    \put(0.6495975,0.08463064){\color[rgb]{0,0,0}\makebox(0,0)[lb]{\smash{$\vdots$}}}%
  \end{picture}%
\endgroup